\documentclass[runningheads]{llncs}
\usepackage{url}
\usepackage[utf8]{inputenc} 
\usepackage[T1]{fontenc}    
\usepackage{hyperref}       
\usepackage{url}            
\usepackage{booktabs}       
\usepackage{amsfonts}       
\usepackage{nicefrac}       
\usepackage{microtype}      
 \usepackage[linesnumbered,ruled]{algorithm2e}
\usepackage{algorithmic}
\usepackage{xcolor}
\usepackage{graphicx}
\usepackage{cite}
\usepackage[numbers]{natbib}

\usepackage{amsmath} 
\AtBeginDocument{%
\providecommand\BibTeX{{%
        \normalfont B\kern-0.5em{\scshape i\kern-0.25em b}\kern-0.8em\TeX}}}

\DeclareMathAlphabet{\mathpzc}{OT1}{pzc}{m}{it}
\DeclareMathAlphabet\mathbfcal{OMS}{cmsy}{b}{n}

\newtheorem{assumption}{Assumption}

\begin{document}
\title{Lyapunov-Based Reinforcement Learning for\\ Decentralized Multi-Agent Control}
%
%
\author{Qingrui Zhang\inst{1} \and
Hao Dong\inst{2} \and
Wei Pan\inst{3}}
\authorrunning{Q. Zhang, H. Dong, and W. Pan}
%
\institute{School of Aeronautics and Astronautics, Sun Yat-Sen University, Guangzhou, China 
\\  \email{qingrui.zhang@tudelft.nl}  \and
Center on Frontiers of Computing Studies, Peking University, Beijing, China
 \\\email{hao.dong@pku.edu.cn} \and
Department of Cognitive Robotics, Delft University of Technology, Delft, the Netherlands\\
 \email{wei.pan@tudelft.nl}}
\maketitle              
\begin{abstract}
    Decentralized multi-agent control has broad applications, ranging from multi-robot cooperation to distributed sensor networks. In decentralized multi-agent control, systems are complex with unknown or highly uncertain dynamics, where traditional model-based control methods can hardly be applied. Compared with model-based control in control theory, deep reinforcement learning (DRL) is promising to learn the controller/policy from data without the knowing system dynamics. However, to directly apply DRL to decentralized multi-agent control is challenging, as interactions among agents make the learning environment non-stationary. More importantly, the existing multi-agent reinforcement learning (MARL) algorithms cannot ensure the closed-loop stability of a multi-agent system from a control-theoretic perspective, so the learned control polices are highly possible to generate abnormal or dangerous behaviors in real applications. Hence, without stability guarantee, the application of the existing MARL algorithms to real multi-agent systems is of great concern, e.g., UAVs, robots, and power systems, etc. In this paper, we aim to propose a new MARL algorithm for decentralized multi-agent control with a stability guarantee. The new MARL algorithm, termed as a multi-agent soft-actor critic (MASAC), is proposed under the well-known framework of  ``centralized-training-with-decentralized-execution''. The closed-loop stability is guaranteed by the introduction of a stability constraint during the policy improvement in our MASAC algorithm. The stability constraint is designed based on Lyapunov's method in control theory. To demonstrate the effectiveness, we present a multi-agent navigation example to show the efficiency of the proposed MASAC algorithm.

\end{abstract}
\section{Introduction}
\label{Sec:Introduction}
Multi-agent system control has intrigued researchers from both
industrial and academic communities for decades,  due to its prospect in broad applications, such as formation flight of unmanned aerial vehicles (UAVs) \cite{Zhang2018AESCTE,Zhang2018TIE},  coordination of multi-robots \cite{Berg2008ICRA,Rezaee2014TIE}, flocking/swarm control \cite{Olfati-Saber2007IEEE,Vasarhelyi2018SR}, distributed sensor networks \cite{Olfati-Saber2005CDC},  large-scale power systems \cite{Guo2000Auto},  traffic and transportation systems \cite{Burmeister1997IEE}, etc. Control of a multi-agent system can be achieved in either a centralized or a decentralized manner. However, a multi-agent system with many subsystems has high state and action dimensions that will dramatically increase the design complexity and computational burdens of a single centralized controller \cite{Bakule2008ARC}. In many applications,  every agent of a multi-agent system only has local control capability with access to local observations, e.g., the cooperation of multiple vehicles \cite{Feddema2002TRA}.  The lack of global control capability and information excludes the possibility of centralized control. Besides, centralized control tends to be less reliable. If the central controller fails, the entire system will break down. As an alternative, decentralized control is capable of handling all the above issues.

Decentralized multi-agent control has been extensively studied \cite{Ren2004JGCD,Ren2009IJC,Wang2013TCST,Cheng2015TCSI}. With the assumption that the agents' dynamics are known and linear, many model-based control algorithms have been proposed for different tasks \cite{Wang2013TCST,Rezaee2014TIE,Cheng2015TCSI}. In control theory, those model-based algorithms can ensure closed-loop stability if a multi-agent system satisfies all the assumptions. The state trajectories of a multi-agent system under a model-based control algorithm will always stay close to or even converge to an equilibrium point \cite{Khalil2002Book}.
However, in most applications, agent dynamics are nonlinear, complicated, and highly uncertain, e.g., robotic systems, UAVs, and power systems. Assumptions made by model-based control algorithms can be barely satisfied in real life. Therefore, model-based control algorithms are restrictive, though theoretically sound.  

Compared with model-based control, deep reinforcement learning (DRL) is more promising for the decentralized multi-agent control for complicated nonlinear dynamical systems, as it can learn controller/policy from samples without using much model information \cite{Sutton2018MIT, Levine2018RL_Tut&Revi, Chen2016ICRA, Chen2017IROS, Everett2019arXiv, Zhang2020MRRL_arXiv, Zhang2020CollisionAvoidance}. Recently, deep RL has obtained significant success in applying to a variety of complex single-agent control problems \cite{Haarnoja2018SAC,Kalashnikov2018QT-Opt,Andrychowicz2020IJRR,Kiran2020AD_Survey}. However, it is more challenging to apply deep RL to decentralized multi-agent control. In multi-agent reinforcement learning (MARL), agents seek the best responses to other agents' policies. The policy update of an agent will affect the learning targets of other agents.  Such interactions among agents make MARL training non-stationary, thus influencing the learning convergence. To resolve the non-stationary issue, a ``centralized-training-with-decentralized-execution'' mechanism was employed, based on which a number of MARL algorithms have been proposed, e.g., MADDPG \cite{Lowe2018Coop_Comp}, COMA \cite{Foerster2017Counterfactual}, mean-field MARL \cite{Yang2018MeanField_DMARL}, MATD3 \cite{Ackermann2019NIPS}, and MAAC \cite{Iqbal2019ArXiv}, etc. Unfortunately, the existing MARL algorithms can not ensure the closed-loop stability for a multi-agent system, while stability is the foremost concern for the control of any dynamical systems. It is highly possible that learned control polices will generate abnormal or risky behaviors in real applications. From a control perspective, the learned control policies fail to stabilize a multi-agent system, so they cannot be applied to safety-critical scenarios, e.g., formation flight of UAVs. 

In this paper, we propose MARL algorithms for decentralized multi-agent control with a stability guarantee. A multi-agent soft actor-critic (MASAC) algorithm is developed based on the well-known ``centralized-training-with-decentralized-execution'' scheme. The interactions among agents are characterized using graph theory \cite{Diestel2000Book}. Besides, a stability-related constraint is introduced to the policy improvement to ensure the closed-loop stability of the learned control policies. The stability-related constraint is designed based on the well-known Lyapunov's method in control theory which is a powerful tool for the design of a controller to stabilize the complex nonlinear systems with stability guarantee. \cite{Khalil2002Book}.

\textbf{\emph{Contributions:}} The contributions of this paper can be summarized as follows. 
\begin{enumerate}
  \item For the first time,  a Lyapunov-based multi-agent soft actor-critic algorithm is developed for decentralized control problems based on the ``centralized-training-with-decentralized-execution''  to guarantee the stability of a multi-agent system.
  \item Theoretical analysis is presented on the design of a stability constraint using Lyapunov's method.
\end{enumerate}

\section{Preliminaries} \label{eq:Prelim}

\subsection{Networked Markov game} \label{subsec:NetworkAgent}

Interactions among $N$ agents are characterized using an undirected graph $\mathcal{G}=\big\langle\mathcal{I},\mathcal{E}\big\rangle$, where $\mathcal{I}:=\left\{1,\ldots,N \right\}$ represents the set of $N$ agents and $\mathcal{E}\subseteq \mathcal{I}\times \mathcal{I}$ denotes the interactions among agents. If an agent $i$ is able to interact with an agent $j$ with $j\neq i$ and $i$,$j\in \mathcal{I}$ , there exists an edge $\left(i\text{, } j\right)\in\mathcal{E}$, and agent $j$ is called a neighbor of agent $i$. For an undirected graph,  $\left(j\text{, } i\right)\in\mathcal{E}$ if $\left(i\text{, } j\right)\in\mathcal{E}$.  The neighborhood of agent $i$ is denoted by $\mathcal{N}_i:=\left\{\forall j\in \mathcal{I} \vert \left(i\text{, } j\right)\in\mathcal{E}\right\}$.  Assume the undirected graph is fully connected, so there exists a path from each node $i\in\mathcal{I}$ to any other nodes $j\in\mathcal{I}$ \cite{Diestel2000Book, Godsil2001Book}. If an undirected graph is strongly connected, information could eventually be shared among all agents via the communication graph.

 A networked Markov game with $N$ agents is denoted by a tuple, $\mathcal{MG}:=\big\langle \mathcal{G}, \mathcal{S},\mathcal{A},\mathcal{P},{r},\gamma\big\rangle$, where $\mathcal{G}:=\big\langle\mathcal{I},\mathcal{E}\big\rangle$ is the communication graph among $N$ agents, $\mathcal{S}:=\bigcup_{i=1}^{N} \mathcal{S}_{i}$ is the entire environment space with $\mathcal{S}_{i}$ the local state space for agent $i\in\mathcal{I}$, $\mathcal{A}:=\bigcup_{i=1}^{N} \mathcal{A}_{i}$ denotes the joint action space with $\mathcal{A}_{i}$ the local action space for agent $i\in\mathcal{I}$, $\mathcal{P}:=\mathcal{S}\times\mathcal{A}\times\mathcal{S}\rightarrow\mathbb{R}$ specifies the state transition probability function, and $r:=\mathcal{S}\times\mathcal{A}\rightarrow \mathbb{R}$ represents the global reward function of the entire multi-agent system.  The global transition probability can, therefore, be denoted by $\mathcal{P}\left(\boldsymbol{s}_{t+1}\vert\boldsymbol{s}_{t}, \boldsymbol{a}_t\right)$. The joint action of $N$ agents is $\boldsymbol{a}=\left\{a_1,\ldots,a_N\right\}$ where $a_i$ denotes the action of an agent $i\in\mathcal{I}$. Accordingly,  the joint policy is defined to be $\boldsymbol{\pi}=\left\{\pi_1,\ldots,\pi_N\right\}$ where $\pi_i$ ($\forall i\in\mathcal{I}$) are local policies for an agent $i$. Hence, the global policy for the entire multi-agent system is defined to be $\pi\left(\boldsymbol{a}\vert \boldsymbol{s}\right) =\prod_{i=1}^N\pi_{i}\left(a_i\vert s_i\right)$. Assume each agent $i$ can only obtain a local observation $s_i\in\mathcal{S}_i$ (e.g. its own states and state information of its neighbors) to make decisions at the execution.

For any given initial global state $\boldsymbol{s}_0$, the global expected discounted return following a joint policy $\boldsymbol{\pi}$ is given by
\begin{equation}
    V\left(\boldsymbol{s}_t\right) = \sum_{t=0}^{\infty}\mathbb{E}_{\left(\boldsymbol{s}_t, \boldsymbol{a}_t\right)\sim\boldsymbol{\rho}_{\boldsymbol{\pi}}}\left[ \gamma^{t} r\left(\boldsymbol{s}_t,\boldsymbol{a}_t\right)\vert \boldsymbol{s}_0\right] \label{eq:GlobalV}
\end{equation}
where $\gamma$ is a discount factor, $V$ is the global value function and $\boldsymbol{\rho}_{\boldsymbol{\pi}}$ is the state-action marginals of the trajectory distribution induced by a global policy $\boldsymbol{\pi}$. The global action-value function (a.k.a. Q-function) of the entire system is 
\begin{equation}
    Q\left(\boldsymbol{s}_t,\boldsymbol{a}_t\right) = r\left(\boldsymbol{s}_t,\boldsymbol{a}_t\right)+\gamma\mathbb{E}_{\boldsymbol{s}_{t+1}}\left[V\left(\boldsymbol{s}_{t+1}\right)\right] \label{eq:GlobalQ}
\end{equation}

\subsection{Soft actor-critic algorithm}
In this paper, the soft actor-critic (SAC) algorithm will be used for the design of the multi-agent reinforcement learning algorithm. The SAC algorithm belongs to off-policy RL that is more sample efficient than on-policy RL methods \cite{Sutton2018MIT}, such as the trust region policy optimization (TRPO) \cite{Schulman2015TRPO} and the proximal policy optimization (PPO) \cite{Schulman2017PPO}.
In SAC, an expected entropy of the policy $\boldsymbol{\pi}$ is added to the value functions (\ref{eq:GlobalV}) and (\ref{eq:GlobalQ}) to regulate the exploration performance at the training stage \cite{Ziebart2010, Haarnoja2018SAC1}. The inclusion of the entropy term makes the SAC algorithm exceed both the efficiency and final performance of  deep deterministic policy gradient (DDPG) \cite{Lillicrap2015DDPG,Haarnoja2018SAC2}. 
With the inclusion of the expected entropy, the action value function (\ref{eq:GlobalV}) to be maximized for a multi-agent system will turn into 
\begin{equation}
     V\left(\boldsymbol{s}_t\right) = \sum_{t=0}^{\infty}\mathbb{E}_{\left(\boldsymbol{s}_t, \boldsymbol{a}_t\right)\sim\boldsymbol{\rho}_{\boldsymbol{\pi}}}\left[ \gamma^{t} \left(r\left(\boldsymbol{s}_t,\boldsymbol{a}_t\right)+\alpha \mathcal{H}\left(\boldsymbol{\pi}\left(\cdot\vert \boldsymbol{s}_t\right)\right) \right)\vert \boldsymbol{s}_0\right] \label{eq:GlobalV_Entropy}
\end{equation}
where $\alpha$ is the temperature parameter used to control the stochasticity of the policy by regulating  the relative importance of the entropy term against the reward, and $\mathcal{H}\left(\boldsymbol{\pi}\left(\cdot\vert \boldsymbol{s}_t\right)\right)=-\mathbb{E}_{\boldsymbol{\pi}}\big[\log\left(\boldsymbol{\pi}\left(\cdot\vert \boldsymbol{s}_t\right)\right)\big]$ is the entropy of the policy $\boldsymbol{\pi}$.  Accordingly, a modified Bellman backup operator is defined as 
\begin{equation}
    \mathcal{T}^{\boldsymbol{\pi}}Q\left(\boldsymbol{s}_t,\boldsymbol{a}_t\right) = r\left(\boldsymbol{s}_t,\boldsymbol{a}_t\right)+\gamma\mathbb{E}_{\boldsymbol{s}_{t+1}}\left[V\left(\boldsymbol{s}_{t+1}\right)\right] \label{eq:Bellma_Op}
\end{equation}
where $V\left(\boldsymbol{s}_{t}\right)=\mathbb{E}_{\boldsymbol{a}_t\sim\boldsymbol{\pi}}\left[ Q\left(\boldsymbol{s}_t,\boldsymbol{a}_t\right)-\alpha \log\left(\boldsymbol{\pi}\left(\boldsymbol{a}_t\vert \boldsymbol{s}_t\right)\right) \right]$.

\subsection{Lyapunov stability in control theory}
A dynamical system is called stable, if its state trajectory starting in vicinity to an equilibrium point will stay near the equilibrium point all the time. Stability is a crucial concept for the control and safety of any dynamical systems. Lyapunov stability theory provides a powerful means of stabilizing unstable dynamical systems using feedback control. The idea is to select a suitable Lyapunov function and force it to decrease along the trajectories of the system.  The resulting system will eventually converge to its equilibrium.  Lyapunov stability of dynamic systems at a fixed policy $\pi\left(a\vert s\right)$ is given by Lemma \ref{lem:LyapStab}.
\begin{lemma}[Lyapunov stability] \label{lem:LyapStab} \cite{Khalil2002Book} Suppose a system is denoted by a nonlinear mapping $s_{t+1}=f\left(s_{t},\pi\left(a_t\vert s_t\right)\right)$. Let $L\left(s_t\right)$ be a continuous function such that $L\left(s_T\right)=0$, $L\left(s_t\right)>0$ ($\forall s_t\in\Omega\; \& \; s_t\neq s_T$), and $L\left(s_{t+1}\right)-L\left(s_{t}\right)\leq 0$ ($\forall s_t\in\Omega$) where $s_T$ is an equilibrium state and $\Omega$ is a compact state space. Then the system $s_{t+1}=f\left(s_{t},\pi\left(a_t\vert s_t\right)\right)$ is stable around $s_T$ and $L\left(s_t\right)$ is a Lyapunov function. Furthermore, if  $L\left(s_{t+1}\right)-L\left(s_{t}\right)<0$ ($\forall s_t\in\Omega$), the system is asymptotically stable around $s_T$.
\end{lemma}
Note that maximizing the objective function (\ref{eq:GlobalV}) or (\ref{eq:GlobalQ}) doesn't necessarily result in a policy stabilizing a dynamical system.

\section{Multi-agent reinforcement learning with Lyapunov stability constraint}
In this section, we will first develop a MARL algorithm based on the SAC algorithm by following a similar idea as the multi-agent deterministic policy gradient descent (MADDPG) \cite{Lowe2018Coop_Comp}. The proposed algorithm is termed as Multi-Agent Soft Actor-Critic algorithm (MASAC). The proposed MASAC algorithm is thereafter enhanced by incorporating a carefully designed Lyapunov constraint.

\subsection{Multi-agent soft actor-critic algorithm} \label{subsec: MASAC}
\begin{figure}[tbph]
    \centering
    \includegraphics[width=0.6\textwidth]{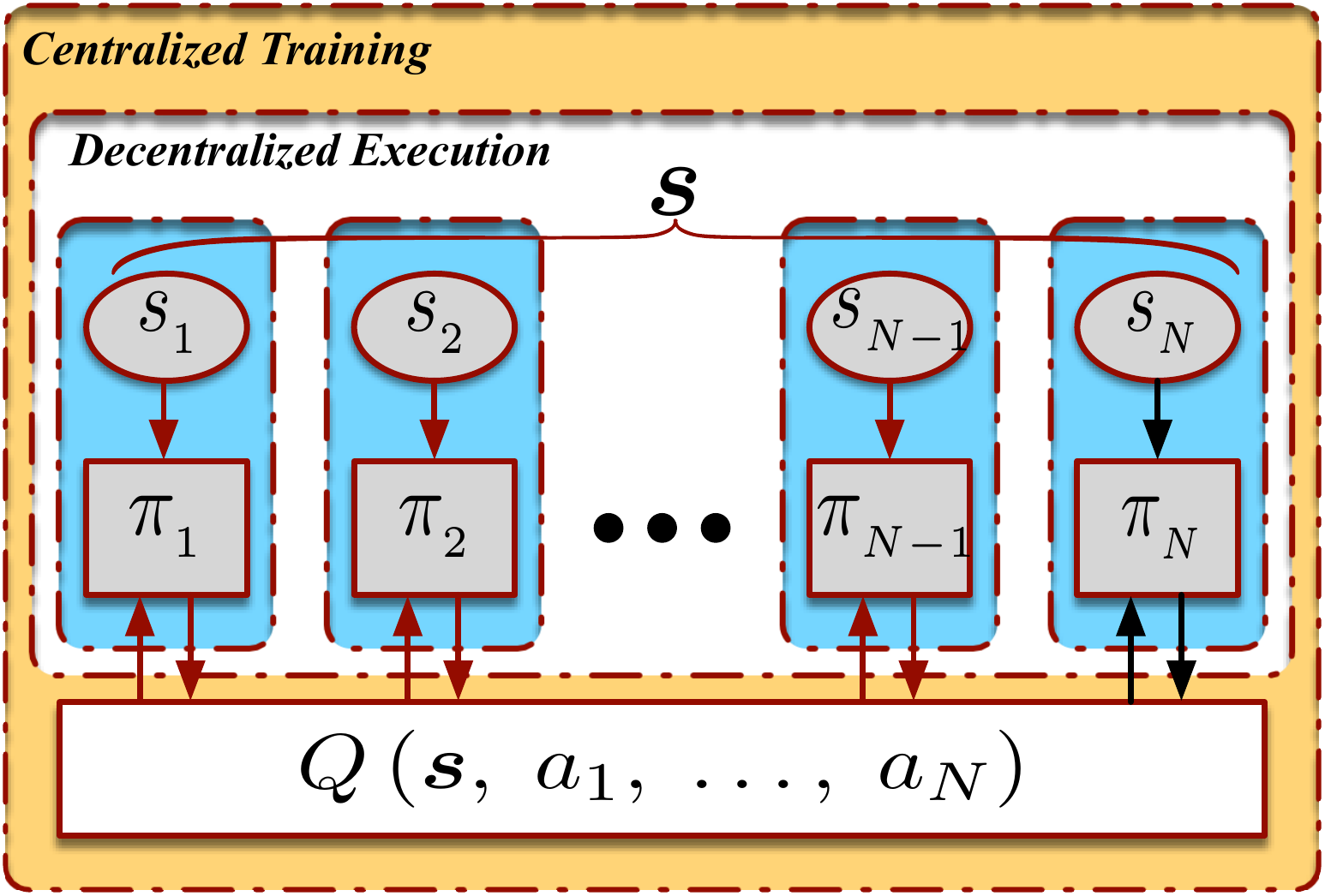}
    \caption{Centralized training with decentralized execution}
    \label{fig:CenTrain_DecExecution}
\end{figure}
The crucial concept behind the MASAC is the so-called ``\emph{centralized training with decentralized execution}'' shown in Figure \ref{fig:CenTrain_DecExecution}. A centralized critic using global information is employed at the training stage, while each agent uses their own independent policy taking local observations as inputs. Hence, at the training stage, agents share their rewards with all the other agents for the calculation of the central critic. In MASAC, it is expected to maximize the entropy-regularized objective function introduced in (\ref{eq:GlobalV_Entropy}).

In decentralized control, agents make decisions based on their local observations \cite{Feddema2002TRA, Ren2004JGCD,Ren2009IJC,Keviczky2006Auto, Bakule2008ARC}. Hence, their polices are assumed to be independent of one another, i.e., $\pi\left(\boldsymbol{a}_{t}\vert\boldsymbol{s}_{t}\right)=\prod_{i=1}^N\pi_{i}\left(a_i\vert s_i\right)$. Hence, the entropy of the joint policy $\boldsymbol{\pi}\left(\boldsymbol{a}_{t}\vert\boldsymbol{s}_{t}\right)$ in (\ref{eq:GlobalV_Entropy}) is 
\begin{equation}
    \mathcal{H}\left(\boldsymbol{\pi}\right) = -\sum_{i}^{N}\mathbb{E}_{{\pi}_{i}}\big[\log\left(\pi_{i}\right)\big] = \sum_{i}^{N} \mathcal{H}\left(\pi_i\right)  \label{eq:GlobalEntropy}
\end{equation}
where $\mathcal{H}\left(\pi_i\right)$ represent the entropy of each local policy $\pi_i$. 

The entire algorithm is divided into policy evaluation and policy improvement. In the policy evaluation step, we will repeatedly apply the modified Bellman backup operator (\ref{eq:Bellma_Op}) to the $Q$-value of a fixed joint policy.  Let the centralized Q-value for the multi-agent system be parameterized by $\theta$. The critic neural network parameter $\theta$ is trained to minimize the following Bellman residual.
\begin{align}
       J_{Q}\left(\theta\right) =& \mathbb{E}_{\left(\boldsymbol{s}_{t},\boldsymbol{a}_{t}\right)\sim\mathcal{D}}\Bigg[\frac{1}{2}\Bigg(Q_{\theta}\left(\boldsymbol{s}_{t}, \boldsymbol{a}_{t}\right) -r\left(\boldsymbol{s}_{t}, \boldsymbol{a}_{t}\right) \Bigg.\Bigg. \nonumber\\
       &\Bigg.\left. -\gamma\mathbb{E}_{\boldsymbol{s}_{ t+1}}\bigg[V_{\bar{\theta}}\left(\boldsymbol{s}_{t+1}\right)+\alpha\sum_{i}^{N} \mathcal{H}\left(\pi_{{\phi}_{i}}\right) \bigg]\right)^2\Bigg] 
\end{align}
In the optimization, the value function is replaced by the Q-value function. Therefore, the critic parameters are optimized by stochastic gradient descent as
\begin{equation}
    \nabla_\theta J_{Q}\left(\theta\right)  = \mathbb{E}_{\left(\boldsymbol{s}_{t},\boldsymbol{a}_{t}\right)\sim\mathcal{D}}\Big[\nabla_\theta Q_{\theta}\left(\boldsymbol{s}_{t}, \boldsymbol{a}_{t}\right)\delta_Q\Big]
\end{equation}
where 
\begin{equation}
    \delta_Q=Q_{\theta}\left(\boldsymbol{s}_{t}, \boldsymbol{a}_{t}\right)-r -\gamma Q_{\bar{\theta}}\left(\boldsymbol{s}_{t+1}, \boldsymbol{a}_{t+1}\right)+\gamma\alpha\sum_{i}^{N} \log\pi_{{\phi}_{i}}
\end{equation}

In policy improvement, the policy is updated according to 
\begin{equation}
    \boldsymbol{\pi}* = arg \min_{\boldsymbol{\pi}'\in\Pi} \mathbb{E}_{{\pi}_{i}}\Big[\alpha\sum_{i}^{N}\log\left(\pi_i\right)-Q\left(\boldsymbol{s}_{t}, \boldsymbol{a}_{t}\right)\Big] \label{eq:globalPI}
\end{equation}
where $\boldsymbol{\pi}^*=\left\{\pi_1^*,\;\ldots,\;\pi_N^*\right\}$ is the optimal joint policy. Assume the policy of agent $i$ is parameterized by $\phi_i$, $\forall i=1,\ldots,N$. According to (\ref{eq:globalPI}), the policy parameters $\phi_i$, $\forall i=1,\ldots,N$ are trained to minimize 
\begin{equation}
    J_{\pi}\left(\boldsymbol{\phi}\right)\simeq\mathbb{E}_{\left(\boldsymbol{s}_{t},\boldsymbol{a}_{t}\right)\sim\mathcal{D}}\Big(\mathbb{E}_{\boldsymbol{\pi}_{\boldsymbol{\phi}}}\Big(\alpha\sum_{i}^{N}\log\left(\pi_{\phi_{i}}\right) -Q_{\theta}\left(\boldsymbol{s}_{t}, \boldsymbol{a}_{t}\right))\Big)\Big) \label{eq:PI_Phi_Global}
\end{equation}

where $\boldsymbol{\phi}=\left\{\phi_1,\; \ldots,\; \phi_N\right\}$ and $\boldsymbol{\pi}_{\boldsymbol{\phi}}=\left\{\pi_{\phi_{1}},\; \ldots,\; \pi_{\phi_{N}}\right\}$. In terms of the stochastic gradient descent, each agent's policy parameter $\phi_i$ will be updated according to
\begin{align}
   \nabla_{\phi_i}J_{\pi}\left(\boldsymbol{\phi}\right)\simeq&\mathbb{E}_{\left(\boldsymbol{s}_{t},\boldsymbol{a}_{t}\right)\sim\mathcal{D}}\Big[\Big(\nabla_{a_{i}}\log\pi_{\phi_{i}}-\nabla_{a_{i}}Q_{{\theta}_{i}}\left(\boldsymbol{s}_{t}, a_{t}, \boldsymbol{\bar{a}}_{t}\right)\Big)\nabla_{\phi_i}a_{{\phi}_{i}}\nonumber\\
   &+\nabla_{\phi_i}\log\pi_{\phi_{i}}\Big] 
\end{align}

The temperature parameter $\alpha$ will be updated based on (\ref{eq:Globalalpha_update}).
\begin{equation}
    J_{\alpha_{i}} =\mathbb{E}_{\boldsymbol{\pi}}\left[-\alpha\sum_{i}^{N}\log \pi_i-\alpha\bar{\mathcal{H}}\right] \label{eq:Globalalpha_update}
\end{equation}
The MASAC algorithm is summarized in Algorithm \ref{alg:MASAC}. The final MASAC algorithm uses two soft Q-functions to mitigate the estimation bias in the policy improvement and further increase the algorithm performance \cite{Hasselt2015DDQN,Fujimoto2018TD3,Ackermann2019ReducingOB}. 
\begin{algorithm}[tb]
   \caption{Multi-agent soft actor-critic algorithm} \label{alg:MASAC}
\begin{algorithmic}
   \STATE Initialize parameters $\theta^1$, $\theta^2$ and $\phi_i$ $\forall i\in\mathcal{I}$ 
   \STATE $\bar{\theta}^1\leftarrow\theta^1$, $\bar{\theta}^2\leftarrow\theta^2$, $\mathcal{D}\leftarrow\emptyset$
   \REPEAT
   \FOR{each environment step}
    \STATE $a_{i, t} \sim \pi_{\phi_{i}}\left(a_{i,t}\vert s_{i, t}\right)$, $\forall i\in\mathcal{I}$ 
    \STATE $\boldsymbol{s}_{t+1}\sim \mathcal{P}_i\left(\boldsymbol{s}_{t+1}\vert \boldsymbol{s}_{t}, \boldsymbol{a}_{t}\right)$, where $\boldsymbol{a}_{t}=\left\{a_{1,t},\;\ldots,\;a_{N,t}\right\}$
    \STATE $\mathcal{D}\leftarrow\mathcal{D}\bigcup \left\{\boldsymbol{s}_{t}, \boldsymbol{a}_{t}, r\left(\boldsymbol{s}_{t}, \boldsymbol{a}_{t}, \right), \boldsymbol{s}_{t+1}\right\}$
    \ENDFOR
    \FOR{each gradient update step}
    \STATE Sample a batch of data, $\mathcal{B}$ , from $\mathcal{D}$
   \STATE $\theta^j\leftarrow\theta^j-\iota_Q \nabla_{\theta} J_{Q}\left(\theta^j\right) $,  $j=1$, $2$
   \STATE $\phi_i\leftarrow\phi_i-\iota_\pi \nabla_{\phi_{i}}J_{{\pi}_{i}}\left(\phi_i\right)$, $\forall i\in\mathcal{I}$
   \STATE $\alpha\leftarrow \alpha - \iota_\alpha \nabla_{\alpha}J_{\alpha}\left(\alpha\right)$
   \STATE $\bar{\theta}^j\leftarrow\tau\theta^j+\left(1-\tau\right)\bar{\theta}^j$, $j=1$, $2$
   \ENDFOR
   \UNTIL{convergence}
\end{algorithmic}
\end{algorithm}

\subsection{Lyapunov stability constraint}
Qualitatively, stability implies that the states of a system will be at least bounded and stay close to an equilibrium state for all the time. The existing MARL algorithms, including the proposed MASAC algorithm in Section \ref{subsec: MASAC}, can find an optimal policy that can maximize either state or action-value functions. However, they do not necessarily produce a policy that ensures the stability of a system. In this section, we offer a possible solution to incorporate Lyapunov stability as a constraint in the optimization of MASAC. 

A Lyapunov function candidate can be constructed based on cost functions $c\left(s, \pi\right)\geq0$ with $c\left(s_T, \pi\left(a_T\vert s_T\right)\right)=0$ and $s_T$ the target/equilibrium state \cite{Berkenkamp2017arXiv, Chow2018arXiv}. One possible choice of the Lyapunov function candidate is an accumulated cost in a finite time horizon, e.g., model predictive control (MPC) \cite{Mayne2000Auto}. Before the introduction of the Lyapunov stability constraint, we make several assumptions on both the cost functions to be designed and the system dynamics of interest. 
The assumption on the cost function is given as follows.
\begin{assumption} \label{assump:Lipsch_Cost}
The cost function $c\left(s, \pi\left(\cdot \vert s\right)\right)$ is bounded $\forall s\in \Omega$ and Lipschitz continuous with respect to $s$, namely $\Vert c\left(s_1, \pi\left(\cdot \vert s_1\right)\right)-c\left(s_2, \pi\left(\cdot \vert s_2\right)\right)\Vert_2\leq l_c\Vert s_1 - s_2\Vert_2$ where $l_c>0$ is a Lipschitz constant.
\end{assumption}
Assumption \ref{assump:Lipsch_Cost} could ensure the Lyapunov function candidate to be bounded and Lipschitz continuous for the state $s$, if we choose it to be an accumulated cost in a finite time horizon. Hence, we could further assume the Lyapunov function candidate is Lipschitz continuous with the state $s$ on a compact set $\Omega$ with $l_L$ the Lipchitz constant.

Since we are interested in the decentralized control problem of multiple agents with deterministic dynamics, the following assumption on the physical dynamics is made.
\begin{assumption} \label{assump:Lipsch_Dyn}
Consider a deterministic, discrete-time agent system $s_{t+1}=f\left(s_{t},a_{t}\right)$. The nonlinear dynamics $f$ is Lipschitz continuous with respect to $a_{t}$, namely $\Vert f\left(s_{t},a_{t}^2\right)-f\left(s_{t},a_{t}^1\right)\Vert_2\leq l_f\Vert a_{t}^2-a_{t}^1\Vert_2$ where $l_f>0$ is a Lipschitz constant.
\end{assumption}
According to the existence and uniqueness theorem, the local Lipschitz condition is a common assumption for deterministic continuous systems.

According to Lemma \ref{lem:LyapStab}, the state $s_{t+1}$ is needed to evaluate the stability of a system under a fixed policy, but $s_{t+1}$ is not available in general. In Theorem \ref{thm:StabEval}, we show that it is possible to evaluate the stability of a new policy $\pi_{new}$, if we already have a feasible policy $\pi_{old}$ associated with a Lyapunov function $L\left(s\right)$. Here, a feasible policy implies that a system is stable and that a Lyapunov function exists.

\begin{theorem} \label{thm:StabEval} 
Consider a system $s_{t+1}=f\left(s_{t},a_{t}\right)$ 
Suppose Assumptions \ref{assump:Lipsch_Cost} and \ref{assump:Lipsch_Dyn} hold. Let $\pi_{old}$ be a feasible policy for data collection and $L_{{\pi_{old}}}\left(s\right)$ is the Lyapunov function. A new policy $\pi_{new}$ will also be a feasible policy under which the system is stable, if there exists
\begin{equation}
    L_{{\pi_{old}}}\left(s_{t+1}\right) + l_Ll_f\Vert a_t^{\pi_{new}}-a_t^{\pi_{old}}\Vert_2 -L_{{\pi_{old}}}\left(s_{t}\right)\leq 0  \label{eq:StabEvalCond}
\end{equation}
where $\forall s_{t}\in \Omega$, $\left(s_t, a_t^{\pi_{old}}, s_{t+1}\right)$ is a tuple from the policy $\pi_{old}$, $l_L$ and $l_f$ are Lipschitz constants of the  Lyapunov function and system dynamics, respectively.
\end{theorem}

\begin{figure}[tbph]
    \centering
     \includegraphics[width=0.85\textwidth]{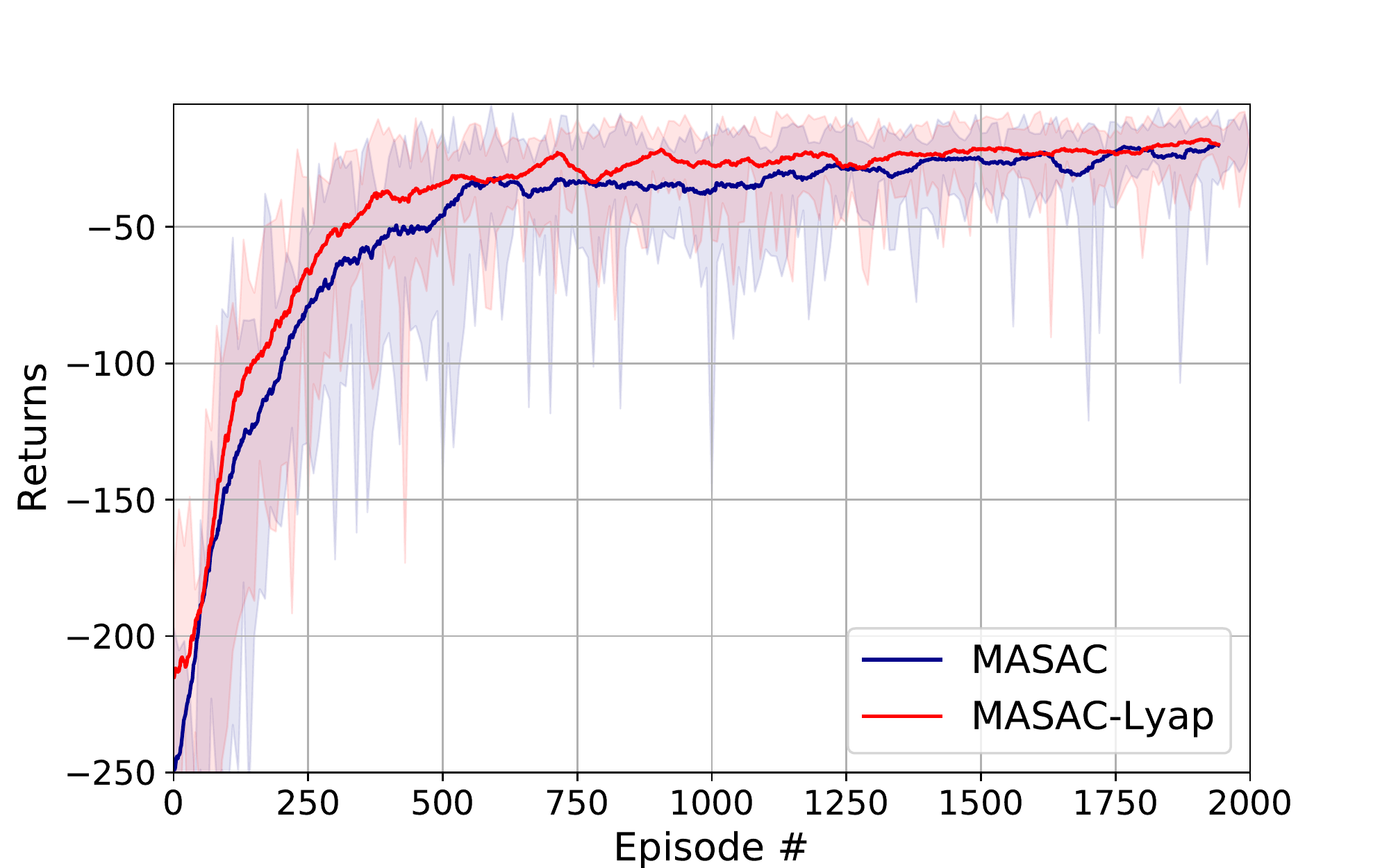}
    \caption{Learning curves of the rendezvous experiment ($40$ steps per episode)}
    \label{fig:Learncurve_Rendezvous}
\end{figure} 

Theorem \ref{thm:StabEval} requires all the states need to be visited to evaluate the stability of a new policy. Unfortunately, it is impossible to visit an infinite number of states. However, Theorem \ref{thm:StabEval} still shows a potential way to use historical samples for the old policies to evaluate the current policy. Based on Theorem \ref{thm:StabEval}, we are able to add a Lyapunov constraint similar to (\ref{eq:StabEvalCond}) in the policy gradient of each agent for DC of multi-agent systems. With the inclusion of the Lyapunov constraint (\ref{eq:StabEvalCond}) \cite{Han2019arXiv, Berkenkamp2017arXiv}, the objective function (\ref{eq:PI_Phi_Global}) is rewritten as 
\begin{equation}
    J_{\pi}\left(\boldsymbol{\phi}\right)\simeq\mathbb{E}_{\left(\boldsymbol{s}_{t},\boldsymbol{a}_{t}\right)\sim\mathcal{D}}\Big(\mathbb{E}_{\boldsymbol{\pi}_{\boldsymbol{\phi}}}\Big(\alpha\sum_{i}^{N}\log\left(\pi_{\phi_{i}}\right) -Q_{\theta}\left(\boldsymbol{s}_{t}, \boldsymbol{a}_{t}\right))+\beta\Delta L_{\boldsymbol{\phi}}\Big)\Big) \label{eq:PI_Phi_Lyap}
\end{equation}
where $\beta\in\left[0, 1\right]$ and $\Delta L_{\psi}=L_i\left(\boldsymbol{s}_{t+1}, \boldsymbol{a}_{t+1}\right)+l_Ll_f\Vert \boldsymbol{a}_{\phi}-\boldsymbol{a}_{t}\Vert_2-L_i\left(\boldsymbol{s}_{t}, \boldsymbol{a}_{t}\right)+\beta c_{\boldsymbol{\pi}}\left(\boldsymbol{s}_{t}\right)$.

At training, the Lyapunov functions $L\left(\boldsymbol{s}_{t}\right)$ will be parameterized by $\psi$ which is trained to minimize
\begin{equation*}
    J_{{L}}\left(\psi\right)=\mathbb{E}_{\left(\boldsymbol{s}_{t},\boldsymbol{a}_{t}\right)\sim\mathcal{D}}\left[\frac{1}{2}\left(L_{\psi}\left(\boldsymbol{s}_{t}, \boldsymbol{a}_{t}\right)-L_{target}\right)^2\right]
\end{equation*}
where $L_{target}=\sum_{t=0}^{T}c\left(\boldsymbol{s}_{t}, \boldsymbol{a}_{t}\right)$ with $T$ denoting a finite time horizon as in model predictive control. The modified robust multi-agent reinforcement learning algorithm is summarized in Algorithm \ref{alg:DecMASAC_Lyap}.
\begin{algorithm}[tbph]
   \caption{Multi-agent soft actor-critic algorithm with a Lyapunov constraint} \label{alg:DecMASAC_Lyap}
\begin{algorithmic}
   \STATE Initialize parameters $\theta^1$, $\theta^2$ and $\phi_i$ $\forall i\in\mathcal{I}$ 
   \STATE $\bar{\theta}^1\leftarrow\theta^1$, $\bar{\theta}^2\leftarrow\theta^2$, $\mathcal{D}\leftarrow\emptyset$
   \REPEAT
   \FOR{each environment step}
    \STATE $a_{i, t} \sim \pi_{\phi_{i}}\left(a_{i,t}\vert s_{i, t}\right)$, $\forall i\in\mathcal{I}$ 
    \STATE $\boldsymbol{s}_{t+1}\sim \mathcal{P}_i\left(\boldsymbol{s}_{ t+1}\vert \boldsymbol{s}_{t}, \boldsymbol{a}_{t}\right)$, where $\boldsymbol{a}_{t}=\left\{a_{1,t},\;\ldots,\;a_{N,t}\right\}$
    \STATE $\mathcal{D}\leftarrow\mathcal{D}\bigcup \left\{\boldsymbol{s}_{ t}, \boldsymbol{a}_{t}, r\left(\boldsymbol{s}_{t}, \boldsymbol{a}_{t}\right), \boldsymbol{s}_{t+1}\right\}$
    \ENDFOR
    \FOR{each gradient update step}
    \STATE Sample a batch of data, $\mathcal{B}$ , from $\mathcal{D}$
   \STATE $\theta^j\leftarrow\theta^j-\iota_Q \nabla_{\theta_{i}} J_{Q_{i}}\left(\theta^j\right) $,  $j=1$, $2$
   \STATE $\phi_i\leftarrow\phi_i-\iota_\pi \nabla_{\phi_{i}}J_{\boldsymbol{\pi}}\left(\boldsymbol{\phi}\right)$, $\forall i\in\mathcal{I}$ from (\ref{eq:PI_Phi_Lyap})
   \STATE $\alpha\leftarrow \alpha - \iota_\alpha \nabla_{\alpha}J_{{\alpha}}\left(\alpha\right)$
   \STATE $\bar{\theta}^j\leftarrow\tau\theta^j+\left(1-\tau\right)\bar{\theta}^j$,  $j=1$, $2$
   \STATE $\psi\leftarrow \psi - \iota_L \nabla_{\psi}J_{L}\left(\psi\right)$
   \ENDFOR
   \UNTIL{convergence}
\end{algorithmic}
\end{algorithm}

\section{Experiment} \label{sec:Experiments}
In this section, we will evaluate our proposed algorithms in a well-known application of multi-agent systems called  ``rendezvous'' \cite{Lin2003CDC}. In the ``rendezvous'' problem, all agents starting from different locations are required to meet at the same target location in the end. Only a subgroup of agents, which are called leaders, have access to the target location, while others need to learn to cooperate with others. In the experiments, both the critic and actor are represented suing fully connected multiple-layer perceptrons with two hidden layers. Each hidden Layer has $64$ neurons with the `ReLU' activation function. The learning rate for the actor network is chosen to be $0.0003$, while the learning rate for the critic network is $0.003$. To stabilize the training, learning rates decrease with a certain decay rate ($0.075^{0.0005}$ in the experiments). The Lyapunov neural network is approximated by an MLP with three hidden layers ($64$ neurons for the first two hidden layers, and $16$ neurons for the last hidden layers). The batch size is selected to be $256$. The parameter $\tau$ for soft updates of both actor and critic networks is picked to be $0.005$. The discount factor $\gamma$ is chosen to be $0.95$. 

The environment is built using the multi-agent environment used in \cite{Lowe2018Coop_Comp}. The agent model in the environment in \cite{Lowe2018Coop_Comp} is replaced by a high-order non-holonomic unicycle model which is widely used in robotics navigation \. The agent dynamics are given as follows.
\begin{equation}
\left\{\begin{array}{ccc}
  \dot{x} &=& v\cos{\psi} \\
  \dot{y} &=& v\sin{\psi} \\
  \dot{\psi} &=& \omega \\
  \dot{v} &=& a \\
  \dot{\omega} &=& r \\
\end{array}
\right.
\end{equation}
where $x$ and $y$ are the positions of the agent, $\psi$ is the heading angle, $v$ is the speed, and $\omega$ is the angular rate. The control actions for each agent are $a$ and $r$, respectively.

\begin{figure}[tbph]
    \centering
     \includegraphics[width=0.85\textwidth]{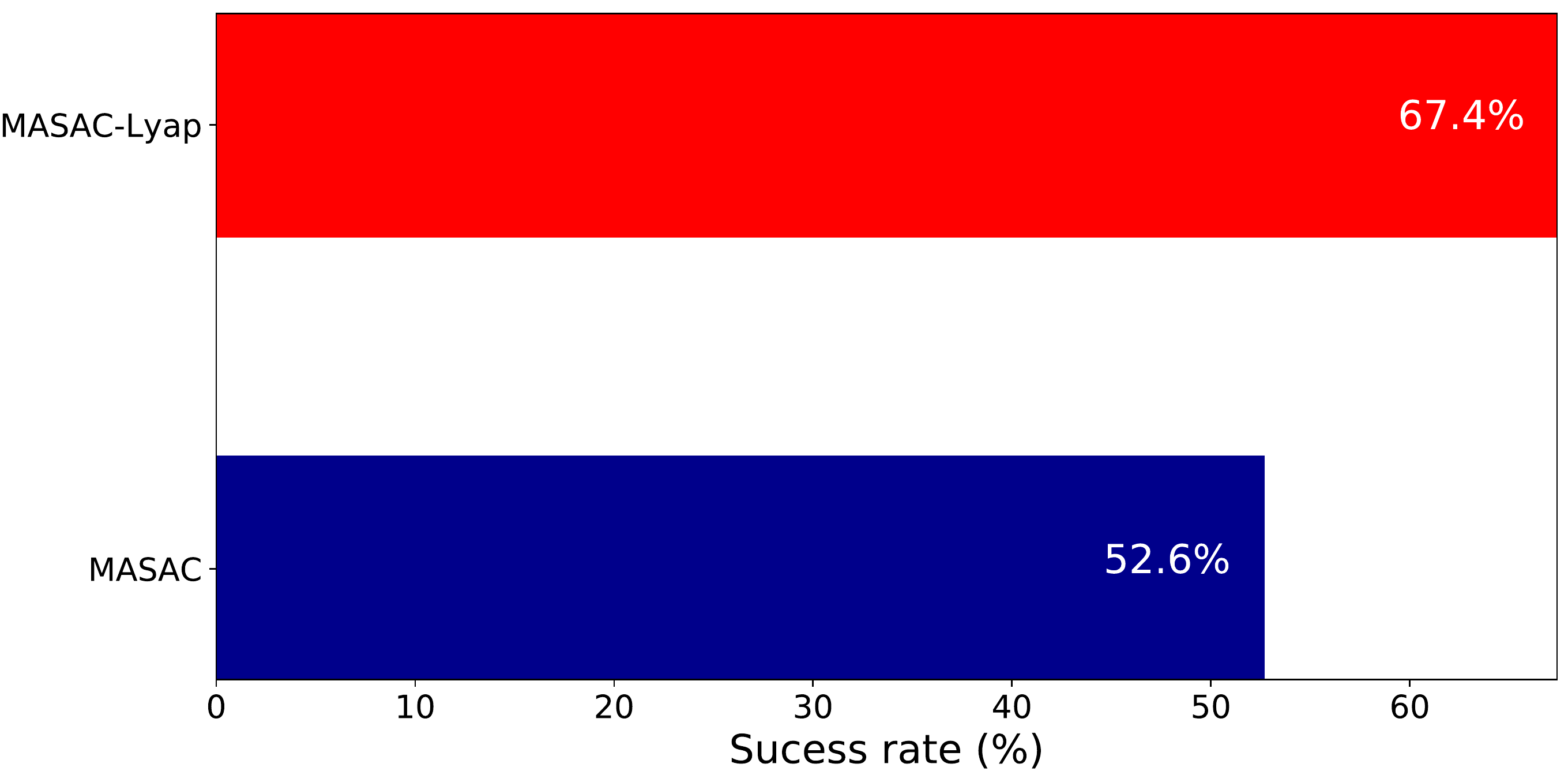}
    \caption{Evaluation results of running the rendezvous experiment for $500$ times using trained polices. ($\text{success rate}=\frac{\text{total number of successful episodes}}{\text{total number of episodes at evaluation}}\times 100$) }
    \label{fig:Evaluation_SuccessRate}
\end{figure}

Learning curves of both the MASAC and MASAC-Lyapunov are shown in Figure \ref{fig:Learncurve_Rendezvous}.  Although both the MASAC and MASAC-Lyapunov will converge, they have a different performance at evaluations.   To further verify the performance, we evaluate both the MASAC and MASAC-Lyapunov for 500 times with agents' initial positions randomly generated. We define a criterion called ``success rate'' to compare the overall performance of the two algorithms. We think an evaluation episode is successful if all agents end up in the target location. The ``success rate'' is calculated by $\frac{\text{total number of successful episodes}}{total number of episodes at evaluation}\times 100$. The ``success rate'' of the two algorithms is shown in Figure \ref{fig:Evaluation_SuccessRate}. With the inclusion of the Lyapunov constraint, we can increase the success rate of the tasks dramatically according to Figure \ref{fig:Evaluation_SuccessRate}. Hence, the Lyapunov constraint will increase the stability performance of the learned policy, thereby resulting in a policy that is more likely to stabilize a system.

\section{Conclusion}
In this paper, we studied MARL for data-driven decentralized control for multi-agent systems. We proposed a MASAC algorithm based on the ``centralized-training-with-decentralized-execution'. We thereafter presented a feasible solution to combine Lyapunov's methods in control theory with MASAC to guarantee stability. The MASAC algorithm was modified accordingly by the introduction of a Lyapunov stability constraint. The experiment conducted in this paper demonstrated that the introduced Lyapunov stability constraint is important to design a policy to achieve better performance than our vanilla MASAC algorithm. 

 \bibliographystyle{splncs04}

\end{document}